# Exciton Transport under Periodic Potential in MoSe$_2$/WSe$_2$ Heterostructures

Zidong Li[1], Xiaobo Lu[2], Darwin F. Cordovilla Leon[3], Jize Hou[1], Yanzhao Lu[1], Austin Kaczmarek[4], Zhengyang Lyu[3], Takashi Taniguchi[5], Kenji Watanabe[5], Liuyan Zhao[3,4], Li Yang[2,6], Parag B. Deotare[1,3]*

[1] Department of Electrical Engineering and Computer Science, University of Michigan, Ann Arbor, MI 48109, USA.
[2] Department of Physics, Washington University in St. Louis, St. Louis, MO 63130, USA.
[3] Applied Physics Program, University of Michigan, Ann Arbor, MI 48109, USA.
[4] Department of Physics, University of Michigan, Ann Arbor, MI 48109, USA.
[5] National Institute for Materials Science, Tsukuba, Japan
[6] Institute of Materials Science and Engineering, Washington University in St. Louis, St. Louis, MO 63130, USA.
*Correspondence: pdeotare@umich.edu

**Abstract:** The predicted formation of moiré superlattices leading to confined excitonic states in heterostructures formed by stacking two lattice mismatched transition metal dichalcogenide (TMD) monolayers was recently experimentally confirmed. Such periodic potential in TMD heterostructure functions as a diffusion barrier that affects the energy transport and dynamics of interlayer excitons (electron and hole spatially concentrated in different monolayers). Understanding the transport of excitons under such condition is essential to establish the material system as a next generation device platform. In this work, we experimentally quantify the diffusion barrier experienced by the interlayer excitons in a hexagonal boron nitride (h-BN) – encapsulated, molybdenum diselenide/tungsten diselenide (MoSe2/WSe2) heterostructure by studying the exciton transport at various temperatures.

TMDs have emerged as a unique class of two dimensional (2D) materials whose low dimensionality has been exploited to reveal, understand and control diverse optical and electrical properties[1–10]. For instance, type-II inorganic heterostructures bonded by van der Waals (vdW) forces can be realized with the use of TMDs[11]. In contrast to traditional semiconductor heterostructures, vdW heterostructures are expected to be free of surface dangling bonds since they do not rely on covalent or ionic bonds, thus preventing elemental diffusion at interface. As a result, this class of heterostructures constitutes an attractive platform for room temperature optoexcitonic devices, which attempt to leverage the enhanced transport properties of the ultralong lived interlayer excitons[12–18]. However, with the recent discovery of localized states at the moiré superlattice's minima, it is crucial to understand the implications of these states on the excitonic properties of heterostructures with different stacking sequences, i.e. H-type (AA' stack) and R-type (AA stack)[19]. Understanding exciton transport under such periodic potential serves as the preamble to establishing TMDs as a universal optoelectronic material platform for next generation devices. By measuring the diffusion of interlayer excitons at various bath temperatures, we observe that the effective diffusivity of excitons can be significantly lower than their intrinsic diffusivity in the absence of the periodic potential. In fact, at low temperatures, most of the excitons remain immobile, ceasing exciton transport completely. Consistent with the expected theoretical values, we estimate an average moiré exciton potential depth of $20\pm4\ meV$ and $34\pm3\ meV$ and extract the in-plane diffusivity in the absence of the barrier to be $0.11\pm0.03\ cm^2/s$ and $0.62\pm0.08\ cm^2/s$ for the H-type and R-type heterostructure respectively.

MoSe$_2$ and WSe$_2$ monolayers were mechanically exfoliated from bulk crystals to fabricate h-BN - encapsulated H-type and R-type heterostructures using a dry transfer process (refer Methods and Supplementary Fig. S1). Fig. 1a shows a bright field image of a H-type sample. The samples for the current work were visually screened for any noticeable defects, followed by confirmation of uniform photoluminescence (PL) intensity over the area of interest to avoid detrimental effects on the excitonic transport properties. For this work, we studied H-type and R-type samples with orientation mismatches of 54° and 5° respectively, for reasons outlined later in the manuscript. A schematic of the moiré superlattices with a twist angle of ~5° is shown in Fig. 1b. The formation of the moiré superlattice was confirmed by the PL signatures at low temperature. Room temperature and low temperature (4 K) PL measurements (2.33 $eV$ continuous-wave excitation, ~1 $\mu m^2$ beam spot size; 5 $nw$ ~ 40 $\mu W$ average power) performed on both H-type and R-type heterostructures are shown in Fig. 1c and d respectively. In both samples, the PL intensity from intralayer excitons was dramatically quenched due to the ultrafast and efficient charge transfer between the individual monolayers (refer Supplementary Fig. S3)[13,14]. For H-type heterostructure, distinct PL emissions peaks from singlet interlayer exciton (peak energy at 1.384 $meV$) and triplet interlayer exciton (peak energy at 1.358 $meV$) with an energy separation of 26 $meV$, corresponding to the splitting of MoSe$_2$'s conduction band due to spin-orbit coupling[14,20–22] were observed, whereas only singlet exciton peak was observed in R-type samples. These transitions are shown in Fig. 1e and f. The broad interlayer exciton peaks at high excitation intensity resolved into narrow peaks with 3 $meV$ linewidth under an ultra-low excitation power (< 20 $nW$), which we attribute to emission from moiré excitons. Consistent with previously reported results[1,12,14,20,21], helicity resolved PL measurements (refer Supplementary Fig. S4) revealed that the moiré exciton states were co-circularly polarized in the R-type heterostructure. On the other hand, for the H-type heterostructure, triplet moiré excitons were co-circularly polarized while singlet moiré excitons were cross-circularly polarized. The helicities of the ascribed moiré exciton emission spectra match the helicities of the interlayer excitons in each sample, supporting the existence of quantized excitonic states within the moiré potential.

The excitons within the moiré potential can be described by an effective Hamiltonian of the form:[2,23]

$$H = \hbar\Omega_0 + \frac{\hbar^2 k^2}{2m^*} + \Delta(r) . \qquad (1)$$

where $\hbar\Omega_0$ is the energy constant that represents the energy of the lowest exciton state without the moiré potential. To emphasize the effect of the moiré potential, we set this reference level to be zero. $m^*$ is effective exciton mass[24] and $\frac{\hbar^2 k^2}{2m^*}$ represents the kinetic energy term. $\Delta(r)$ is the exciton moiré potential, which is determined by the twist angle between the monolayers forming the heterostructure. Fig. 2a, b show a color contour plot representing the exciton moiré potential in H-type and R-type twisted bilayer heterostructures. They are obtained by an interpolation of exciton energies solved by the first-principles simulation of the Bethe-Salpeter Equation (BSE).[25,26] In H-type twisted heterostructures, excitons feel a weak potential barrier, and the highest barrier is up to 30 $meV$. In R-type twisted heterostructures, the variation of the exciton moiré potential is much larger. The barrier height is between 76 $meV$ to 100 $meV$.

By solving this Hamiltonian including first-principles calculated exciton moiré potential with the plane-wave expansion method[23,27,28], the modified exciton energies according to the twisting angle were calculated and are shown in Fig. 2c, d for H-type and R-type twisted heterostructures, respectively. For both twisting types, the energy of the lowest exciton increases with the twist angle. This is owing to the enhancement of the zero-point energy of the quantum well-like moiré potential. Small twist angles lead to a deep and complex energy manifold that can result in the trapping of excitons at the potential minima even at room temperature. Therefore, we chose to work with samples with large orientation mismatch that support only one quantized state within the moiré potential and is comparable to the thermal energy at room temperature.

The diffusion of the interlayer excitons was spatially and temporally monitored by imaging the time-dependent change in exciton density at various temperatures using the technique outlined in previous work[9,29,30]. The normalized exciton density at each time step was fitted with a Gaussian function to extract the temporal evolution of the mean-squared displacement (MSD) of the distribution. The temperature dependent diffusivity ($D$) of interlayer excitons was then estimated from the MSD $\langle(\Delta x(t))^2\rangle$, according to $\langle(\Delta x(t))^2\rangle \equiv \sigma^2(t) = \sigma^2(0) + 2Dt$, where $\sigma$ represents the standard deviation of the Gaussian distribution.

We observed anomalous diffusive transport (nonlinear evolution of the exciton distribution's MSD) in early time regimes, which is attributed to non-equilibrium many-body effects[12,30,31]. Diffusivities of the thermalized interlayer excitons were extracted from the linear MSD evolution regime observed later in time. The evolution of diffusivity with temperature is shown in Fig. 3c We observed near-zero diffusivities at low temperatures, implying the trapping of interlayer excitons at the moiré superlattices. We observed that the average diffusivity of interlayer excitons increased with temperature suggesting the detrapping of interlayer excitons due to thermal energy. The inset in Fig. 3c shows the temperature dependent diffusivity in the low temperature regime. The earlier rise in diffusivity with temperature in the H-type samples indicates a shallower diffusion barrier when compared to the R-type heterostructures. In such a scenario the temperature-activated exciton diffusivity can be modelled as:

$$D = D' \times e^{-\frac{U}{k_B T}} \qquad (2)$$

where $U$ is the diffusion barrier and $D'$ is defined as the in-plane diffusivity in the absence of the barrier. Since we monitor the effective diffusivity at a later time, we neglect density dependent effects such as bandgap renormalization, or screening of the exciton moiré potential[32].

We estimate a diffusion barrier of $20\pm4\ meV$ for the H-type heterostructure and $34\pm3\ meV$ for the R-type heterostructure. We note that the measured value of the barrier represents the spatially averaged diffusion barrier experienced by the excitons. As shown in Fig. 2a, b, the confinement potential is anisotropic, and it is hard to quantitatively derive an effective barrier. As a result, we assumed that the effective barrier to exciton motion should be somewhere between the maximum and minimum energy barriers marked in Fig. 2c, d by the dashed and dotted lines, respectively. The first-principles simulations provide a range of diffusion barriers to compare with the experimentally obtained values. As shown in Fig. 2d, the lowest exciton energy of R-type heterostructures is about $53\ meV$ for a $5°$ twisting angle. Given that the lower and upper bounds of energy barriers (dash and dot lines) are around $76\ meV$ and $100\ meV$, respectively, the effective diffusing barrier lies between $23\ meV$ and $47\ meV$, which agrees with our measured value of $34\ meV$. For the H-type heterostructures, the barrier is shallower. The upper bound of the potential is around $30\ meV$. As shown in Fig. 2c, the energy of the lowest exciton is about $10\ meV$ for a $6°$ twisting angle. Thus, the upper bound of the diffusion barrier is about $20\ meV$, which also quantitively agrees with our measured value of $20\ meV$.

The extracted in-plane diffusivity in the absence of moiré potential was estimated to be $0.11\pm0.03\ cm^2/s$ for H-type and $0.62\pm0.08\ cm^2/s$ for R-type. The difference in $D'$ is attributed to the anisotropic diffusivities of singlet and triplet excitons. In the H-type heterostructure, scattering from singlet to triplet (refer Supplementary Fig. S6) leads to a long-lived triplet state. Given the later time window chosen to analyze the data, we expect diffusion to be dominated by the triplet state in the H-type heterostructure, while diffusion is primarily due to the singlet state in the R-type heterostructure.

The diffusion barrier due to the periodic potential experienced by the excitons should also present its effect of the temporal dynamics of the excitons. It is well known that interlayer excitons are long lived – for periods up to nanoseconds – in contrast to hundreds of picoseconds for intralayer exciton due to the small overlap between the electron and hole wavefunctions[4,11,13,33,34]. We observed ultralong-lived (µs timescale) interlayer excitons at 4 K in the R-type heterostructure. We believe such three orders of magnitude difference arises due to excitons confined in the moiré states. Fig. 4a shows a set of spatially integrated time-resolved photoluminescence (TRPL) measurements for a constant excitation fluence (5 $\mu J/cm^2$) at various temperatures ($T$) ranging from 4 K to room temperature. For $T \geq 120$ K, the TRPL follows a single exponential decay, confirming that the experiments are conducted in low-excitation regime, where exciton-exciton auger recombination can be neglected. At relatively low temperature, the TRPL curves become biexponential. Such behavior is typical of a three-level system as shown in Fig. 4b, where the moiré exciton state is modelled as a long-lived trap state. In this case, the first component depicts thermalization between the moiré state and the interlayer exciton state (approximately equal to $k_{12} + k_{21}$ in Fig. 4b). The second component is the thermal average of the decays from the interlayer exciton state and the moiré state to the ground state. For a fixed energy separation $U$ between the two radiating states in a three-level system, this can be modeled as[35,36]:

$$\tau^{-1} = \left(\frac{e^{-\frac{U}{k_B T}}}{1+e^{-\frac{U}{k_B T}}}\right)\tau_1^{-1} + \left(\frac{1}{1+e^{-\frac{U}{k_B T}}}\right)\tau_2^{-1} \tag{3}$$

where, the temperature dependence of the non-radiative decay rates is modelled by the Boltzmann factor as $\frac{k_{12}}{k_{21}} = e^{-\frac{U}{k_B T}}$. As the temperature increases, the first component gradually disappears since thermalization is achieved at time scales beyond the current experimental limits. In this case, the decay is represented by a single exponential. Fig. 4c shows the extracted lifetime component after thermalization (second component in case of biexponential decay) for various temperatures. We observed that the interlayer excitons are short lived and independent of temperature until the system is cooled below 90 K. A sharp increase in lifetime was observed between 60 K to 20K, beyond which the lifetime plateaued out. Similar temperature-dependent lifetime trends have been reported[35,37–39] in systems whose decay dynamics are described by a three-level system, which consists of two closely energy separated emitting states in thermal equilibrium.

At sufficiently low temperature, where the thermal energy is significantly lower than $U$, a large portion of interlayer excitons remain trapped in the lower moiré state ($k_{12}$ is small), leading to a long lifetime that is dominated by the decay rate $k_1$ of the long-lived moiré state. With increase in temperature, the upper short-lived state is thermally populated, resulting in a sharp drop of observed exciton lifetime as shown in Fig. 4c. A fit to the measured lifetimes with equation (3) is shown with the red solid curve in the figure. The extracted energy separation from the fit was $29 \pm 4\ meV$, in agreement with the value estimated from the transport experiments. However, the same model cannot be applied to H-type heterostructures because of the different energy levels and lifetimes (refer Supplementary Fig. S5, S6) of singlet and triplet states.

In conclusion, we experimentally quantified the diffusion barrier experienced by interlayer excitons in both R-type and H-type $MoSe_2/WSe_2$ heterostructures. Such barrier arises from the periodic moiré potential formed due to the lattice mismatch and the twist angle between monolayers forming the heterostructure. Since the periodicity of the moiré potential and the associated energy of the moiré excitonic states can be controlled by the rotational misalignment between the monolayers, the diffusion barrier can be tuned to control the transport properties. The anisotropy in the periodic potential can be further engineered to combine diverse properties in TMDs to achieve novel functionalities in optoelectronic devices. Together with the enhanced light matter interaction, room temperature stable exciton, valley polarization, and bandgap tunability of TMD's, such heterostructures can emerge as an attractive platform for next generation optoexcitonic devices.

Figures:

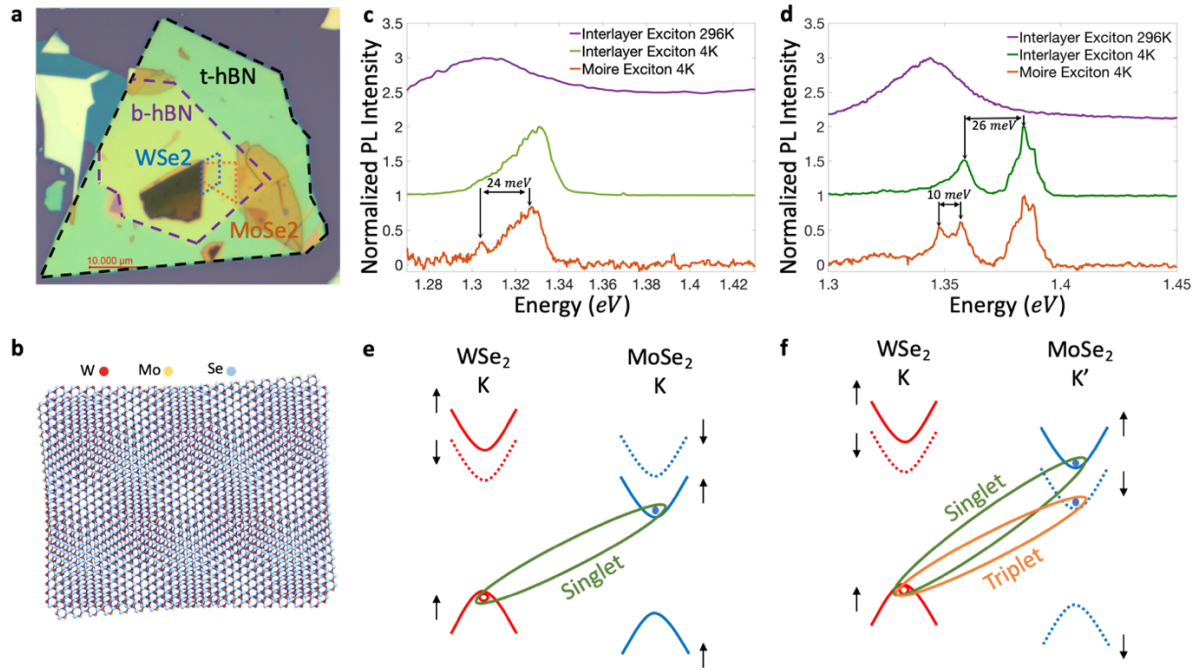

**Fig. 1| Moiré exciton photoluminescence. a**, Bright field optical image of h-BN encapsulated $MoSe_2/WSe_2$ heterostructure. t-hBN (b-hBN) stands for the top (bottom) layer of h-BN. **b**, Schematic of the moiré superlattice formed in a $MoSe_2/WSe_2$ heterostructure with a non-zero twist angle. **c, d**, Comparison of interlayer exciton photoluminescence from a R-type heterostructure (**c**) and an H-type heterostructure (**d**) at room temperature (40 $\mu W$), 4 K using high excitation power (1000 $nW$) and 4 K using low excitation power (20 $nW$). **e, f**, Schematic illustration of singlet and triplet states in a R-type (**e**) and H-type (**f**) heterostructure. Solid (dashed) lines indicated the spin-up (down) state.

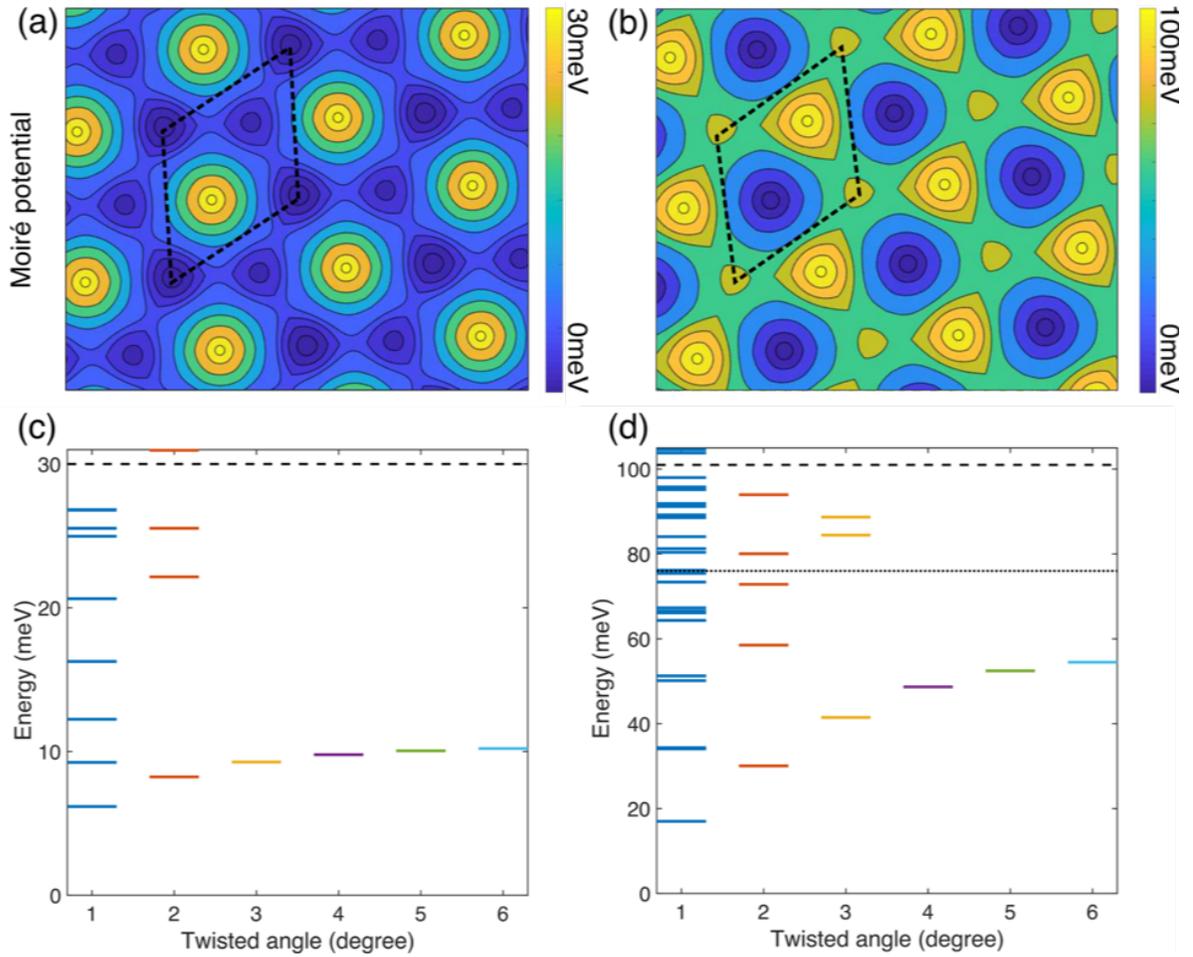

**Fig. 2 | Theoretical predictions of exciton moiré potentials. a,b,** Schematic plot of moiré potential of interlayer exciton of H-type and R-type MoSe$_2$/WSe$_2$ heterostructures, respectively. **c,d,** The evolution of exciton energy according to the twisted angle of H-type and-R type heterostructures, respectively. The energy of the lowest exciton without moiré potential is set to be zero. The upper and lower bounds of moiré potential barrier are indicated by dash and dotted lines, respectively.

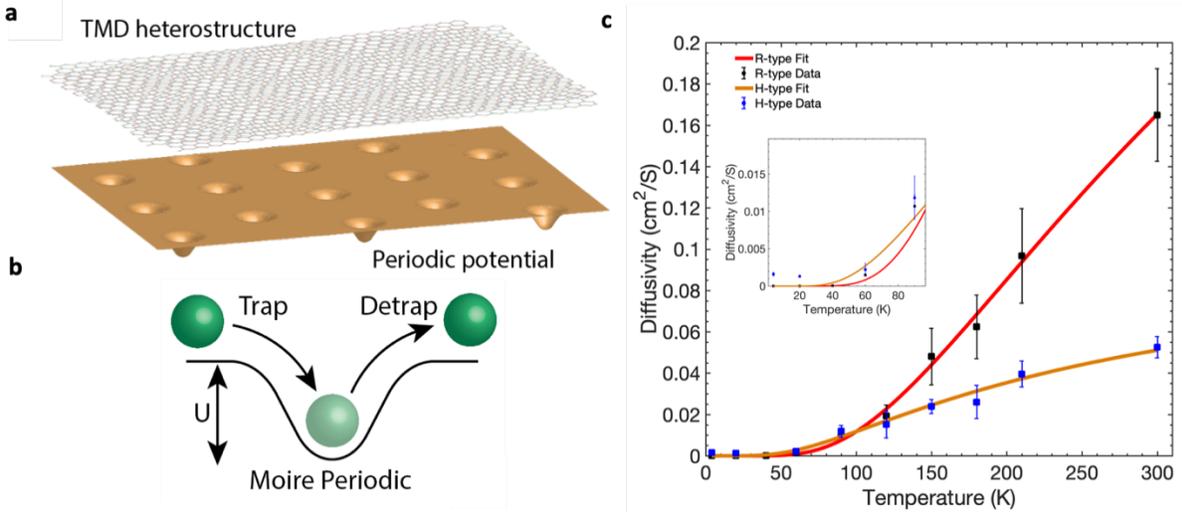

**Fig. 3 | Interlayer exciton diffusion in moiré potentials. a**, Simplified energy landscape of a twisted TMD heterostructure. **b**, Schematic showing the spatially trapping and detrapping process of interlayer excitons. **c**, Temperature dependent diffusivity of interlayer excitons in a R-type (black square) and a H-type (blue square) heterostructure. The solid lines are fits using equation (2). The inset shows low temperature diffusivity regime.

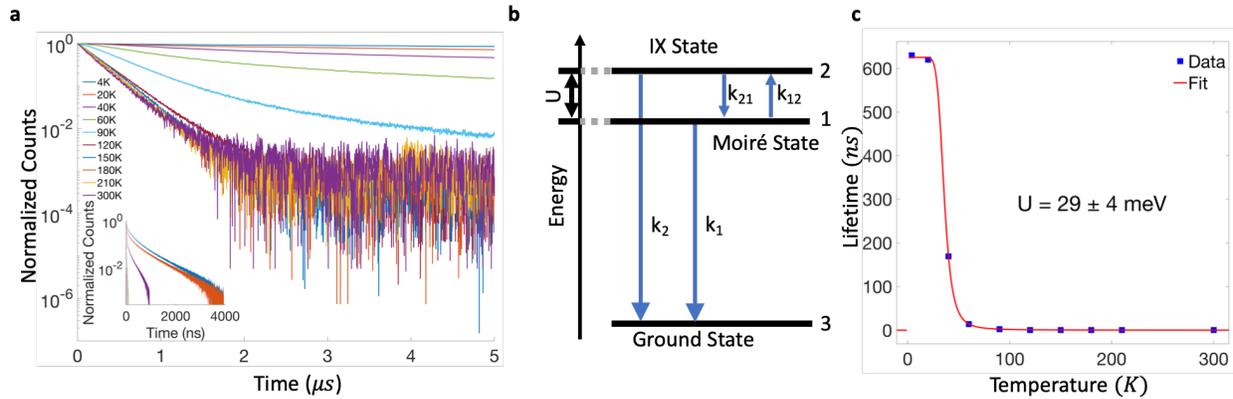

**Fig. 4 | Interlayer exciton decay dynamics in a R-type heterostructure. a**, Temperature dependent Time-resolved photoluminescence (TRPL) of interlayer excitons. The inset shows the ultralong-lived interlayer excitons at low temperature in a 4 $\mu s$ window. **b**, Three level system used to describe the decay processes for interlayer excitons in the R-type heterostructure. $k_1$ ($k_2$) is the decay rate from the moiré state (interlayer exciton state) to ground state. $k_{12}$ and $k_{21}$ are the transition rate between interlayer exciton state and moiré state **c**, Temperature dependence of interlayer exciton lifetime. Lifetime is obtained from the slow component of a bi-exponential fit of the TRPL curves at temperatures below 120 K and from a single-exponential fit of TRPL at temperatures equal or above 120 K. The solid line is a three-level fit using equation (3).